\documentclass{aa}
\usepackage{upgreek}
\usepackage{graphicx}
\usepackage{subcaption}
\usepackage{threeparttable}
\usepackage[varg]{txfonts}
\usepackage{natbib}
\usepackage{float}
\usepackage{soul}
\newlength{\linwx}
\usepackage{color}

\begin{document}

\title{Enriching inner discs and giant planets with heavy elements}

\author{
Bertram Bitsch \inst{1}, \and Jingyi Mah\inst{1}
}
\offprints{B. Bitsch,\\ \email{bitsch@mpia.de}}
\institute{Max-Planck-Institut f\"ur Astronomie, K\"onigstuhl 17, 69117 Heidelberg, Germany
}
\abstract{Giant exoplanets seem to have on average a much higher heavy-element content than the Solar System giants. Past attempts to explain this heavy-element content include collisions between planets, accretion of volatile rich gas, and accretion of gas enriched in micrometre-sized solids. However, these different theories individually could not explain the heavy-element content of giants and the volatile-to-refractory ratios in the  atmospheres of giant planets at the same time. Here we   combine the approaches of gas accretion enhanced with vapour  and small  micrometre-sized dust grains within one model. To this end, we present detailed models of inward-drifting and evaporating pebbles, and describe how these pebbles influence the dust-to-gas ratio and the heavy-element content of the disc. As  pebbles drift inwards, the volatile component evaporates and enriches the disc. At the same time, the smaller silicate core of the pebble continues to move inwards. As the silicate pebbles are presumably smaller than the ice grains, they drift more slowly, leading to a pile-up of material inside of  the water-ice line, increasing the dust-to-gas ratio in this region. Under the assumption that these small dust grains follow the motion of the gas even through the pressure bumps generated by the gaps between planets, gas accreting giants can accrete large fractions of small solids in addition to the volatile vapour. We find that the effectiveness of the solid enrichment requires a large disc radius to maintain the pebble flux for a long time and a high viscosity that reduces the size and inward drift of the small dust grains. However, this process depends crucially on the debated size difference of the pebbles that are inside and outside of the water-ice line. On the other hand, the volatile component released by the inward-drifting pebbles can lead to a high enrichment with heavy-element vapour, independently of a size difference of pebbles inside and   outside the water-ice line. Our model emphasises the importance of the disc's radius and viscosity to the enrichment of dust and vapour. Consequently, we show how our model could explain the heavy-element content of the majority of giant planets by using combined estimates of dust and vapour enrichment.
}

\authorrunning{Bitsch \& Mah}\titlerunning{Enriching inner discs and giant planets with heavy elements}\maketitle

\section{Introduction}
\label{sec:Introduction}

Planet formation theories are classically constrained by the masses, radii, and orbital distances of planets (e.g. \citealt{2004ApJ...604..388I, 2009A&A...501.1139M, 2014A&A...565A..96G, 2017ASSL..445..339B, 2018MNRAS.474..886N}). However, with ever more detailed observations additional constraints on formation theories become available, such as  atmospheric abundances (e.g. \citealt{2019ApJ...887L..20W, 2021Natur.598..580L, 2023arXiv230507753A}) and the total heavy-element content of giant planets \citep{2016ApJ...831...64T, 2023MNRAS.tmp.1812B}, which we focus on in this work.

In order to determine the heavy-element content of giant planets, detailed measurements of the planetary mass and radii are needed. Interior models can then estimate the heavy-element content of gas giants \citep{2016ApJ...831...64T, 2023MNRAS.tmp.1812B}. \citet{2016ApJ...831...64T} determined the heavy-element content of hot Jupiters, finding surprisingly high heavy-element content, where a Jupiter-mass planets should contain on average $57.9\pm 7.0$ Earth masses of heavy elements, nearly twice as much as our own Jupiter (e.g. \citealt{2017GeoRL..44.4649W, 2018A&A...610L..14V}). This high heavy-element content  is difficult to explain just with planetesimal accretion (e.g. \citealt{2020A&A...634A..31V}), where large-scale migration might be helpful \citep{2020A&A...633A..33S}.

Several alternative theories have been proposed to explain the high heavy-element content. \citet{2020MNRAS.498..680G} suggest  that mergers between gas giants could increase their heavy-element content because a large fraction of the gas is lost upon collisions, while their total core masses should be preserved. However, this method requires multiple collisions between giants to explain the most enriched planets, which seems unlikely from an N-body perspective  (e.g. \citealt{2012ApJ...751..119B, 2020A&A...643A..66B}).

Inward-drifting and evaporating pebbles enrich the inner disc with large amounts of volatile (e.g. H$_2$O) vapour (e.g. \citealt{2015ApJ...815..109P, 2017MNRAS.469.3994B, 2021A&A...654A..71S}). Accretion of these large amounts of vapour could naturally explain the heavy-element content of most massive giant planets \citep{2021A&A...654A..71S}, where planets below two Jupiter masses seem to be enriched to higer values \citep{2016ApJ...831...64T} compared to the model. Considering the new interior analysis of \citet{2023MNRAS.tmp.1812B}, on the other hand, gives a better agreement between model predictions on constraints derived from observations.

This idea was also applied to the Solar System \citep{2006MNRAS.367L..47G, 2021A&A...654A..72S}, which can naturally explain the nitrogen content of Jupiter without invoking large-scale migration from the N$_2$ snow line (e.g. \citealt{2015A&A...582A.112B, 2019A&A...632L..11B, 2019AJ....158..194O}). On the other hand, these models cannot naturally explain the refractory component of the Solar System giants \citep{2021A&A...654A..72S} because refractory species cannot be accreted through the vapour phase at large orbital distances. Furthermore, the pebble drift and evaporation scenario is supported by disc observations showing enhancements of the disc's C/H ratio, due to  evaporating pebbles \citep{2020ApJ...891L..16Z, 2020ApJ...903..124B, 2023arXiv230703846B}.

\citet{2023arXiv230601653M} suggest that the enrichment of small dust grains in the inner disc could significantly enhance the local dust-to-gas ratio. This enrichment is based on the different pebble sizes inside and outside the water-ice line, where the small refractory-rich pebbles in the inner disc move more slowly than the larger ice-rich pebbles in the outer disc. Consequently, a `traffic jam' is generated (e.g. \citealt{2017A&A...608A..92D}), enhancing the dust-to-gas ratio in the inner disc. The authors analytically estimate that dust-to-gas ratios of up to 30\% in the inner disc region could be achieved in discs over a long period of time.

This high dust-to-gas ratio also has consequences for the composition of giant planets. \citet{2023arXiv230601653M} argue that growing giant planets in the inner disc region could accrete these small dust grains effectively with the gas, despite the gap opened by the growing giant. Their argument is based on the assumption that grains in the inner disc region are very small and follow the motion of the gas. Consequently, the pressure bump that the giant planets generate beyond their orbits is not effective in blocking the small dust grains, allowing dust accretion onto the planets. This allows gas-accreting planets to accrete large amounts of the small dust grains that follow the motion of the gas. However, they did not model the evolution of dust and gas in detail. 

Here we present a study that takes both the small dust grains and the vapour content of the gas into account by using a viscous disc evolution model combined with a pebble growth and drift model. We then compute the heavy-element content of growing giant planets, allowing a detailed comparison to constraints derived from observations (e.g. \citealt{2016ApJ...831...64T}).

In addition, this approach allows us to understand the contribution of vapour-enhanced gas and small dust grains to the heavy-element content of gas giants, thus also giving insights into  the refractory versus volatile budget of giant planets that is thought to constrain the planet formation pathway (e.g. \citealt{2020A&A...633A..33S, 2021A&A...654A..72S, 2021ApJ...909...40T, 2022arXiv221109080C, 2022A&A...665L...5K}).

\section{Model}
\label{sec:model}

Our model is based on the {\it chemcomp} code outlined in \citet{2021A&A...654A..71S}. This code uses a 1D viscous disc evolution model with the $\alpha$ viscosity \citep{1974MNRAS.168..603L} that allows the viscous evolution of the disc. We include pebble growth and drift following the two-population models of \citet{2012A&A...539A.148B}, as well as evaporation and recondensation of pebbles at evaporation fronts. The effect of the outward-diffusing vapour that recondenses can be seen as spikes in the pebble surface density (and consequently in the dust-to-gas ratio) around ice lines (e.g. \citealt{2021A&A...654A..71S}). Our initial disc set-up is described in Appendix~\ref{ap:parameters}. We use a solar metallicity for our simulations and investigate higher metallicities in Appendix~\ref{ap:metallicity}.

Our model uses initially sub-micrometre-sized grains that then subsequently grow through coagulation and drift inwards. This growth of the dust grains into pebbles scales with the orbital frequency (e.g. \citealt{2008A&A...480..859B}), indicating that grains in the outer disc take longer to grow and drift inwards compared to grains in the inner disc. The sizes of the pebbles are set by the drift limit, which is  the size when pebbles start to drift efficiently,  and the fragmentation limit, which is  the size above which pebbles fragment upon collision (e.g. \citealt{2012A&A...539A.148B}). The fragmentation limit is set by the fragmentation velocity $u_{\rm f}$, which sets the maximum speed particles can have before they fragment upon collision. This size is given as 
\begin{equation}
\label{eq:frag}
 a_{\rm frag} \simeq \frac{2}{3\pi} \frac{\Sigma_{\rm g}}{\rho_{\rm s} \alpha} \frac{u_{\rm f}^2}{c_{\rm s}^2} \ ,
\end{equation}
where $\Sigma_{\rm g}$ is the gas surface density, $\rho_{\rm s}$ the pebble density, and $c_{\rm s}$ the sound speed of the disc. The fragmentation velocity itself is determined in laboratory experiments that yield 1m/s to 10m/s for silicate and water-ice particles, respectively \citep{2015ApJ...798...34G}. These differences in the fragmentation velocity at the water-ice line   then lead to particle size differences of a factor of 100. As larger particles drift more quickly through the disc (e.g. \citealt{2008A&A...480..859B, 2012A&A...539A.148B}), a traffic jam is created inside the water-ice line \citep{2017A&A...608A..92D}, influencing the dust-to-gas ratio of the inner disc. We investigate how this transition affects the composition of the disc below.

However, recent laboratory experiments have cast doubt on this large difference in the fragmentation velocities between silicate and water-ice particles \citep{2019ApJ...873...58M}. Furthermore there is evidence that the  `bouncing barrier' prevents growth above a certain particle size \citep{2010A&A...513A..57Z}, independently of the composition of the particles. We show the behaviour of discs without any transition in the pebble fragmentation velocity in Appendix~\ref{ap:velocity}.

We do not explicitly model the growth and composition of giant planets in this work, but instead follow the approach of \citet{2023arXiv230601653M}. They argue that the heavy-element content of the gas accreted by the planet is set by the dust-to-gas ratio originating from the small dust grains that follow the motion of the gas. This argument is based on the idea that small dust grains follow the motion of the gas and move inwards, even through pressure bumps exerted by growing planets. In addition, we also calculate the vapour component, which is  fueled by evaporating volatile rich pebbles,  of the gas phase. Thus, our estimate of the heavy-element content of giant planets is a sum of the vapour component and the small dust component. However, as we argue below, more detailed calculations are needed to fully understand the heavy-element content of giant planets. Furthermore, outer giant planets could also influence the composition of the inner disc because they block pebbles from reaching the inner disc (e.g. \citealt{2014A&A...572A..35L, 2018arXiv180102341B}), eventually delaying the enrichment of the inner disc with dust (e.g. \citealt{2023arXiv230105505S}) and vapour for a few million years. Furthermore, growing outer planets can take away a significant fraction of the available gas, eventually starving the inner planets \citep{2022arXiv221116239B}, leading to a different planetary growth pathway as well. Our simple approach here   demonstrates the importance of various disc parameters on the heavy-element content (of dust and vapour) in discs, and of growing single giant planets.

\section{Disc enrichment}
\label{sec:disc}

\begin{figure*}
 \centering
 \includegraphics[scale=0.47]{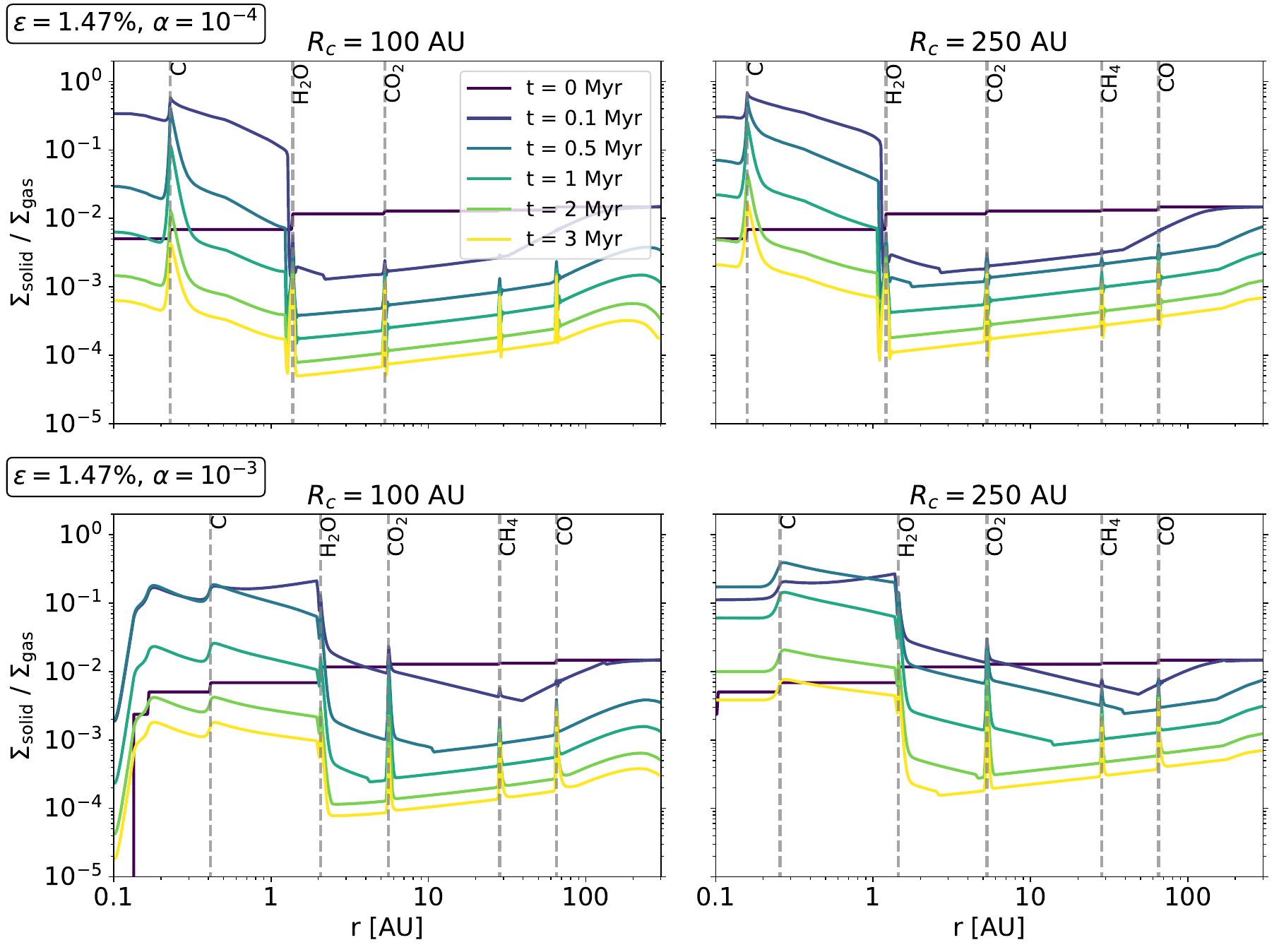}
 \caption{Time evolution of the dust-to-gas ratio in protoplanetary discs with $\alpha = 10^{-4}$ (top) and $\alpha=10^{-3}$ (bottom) and disc radii of $R_{\rm c}=100$ AU (left) and $R_{\rm c}=250$ AU (right). The vertical lines indicate the evaporation fronts of the different chemical species, where inward-drifting pebbles evaporate and recondense, leading to pile-ups in the solid density.
   \label{fig:dtg}
   }
\end{figure*}

Figure~\ref{fig:dtg} shows the dust-to-gas ratio of different simulations as a function of time for discs with a transition in fragmentation velocity of 1 to 10m/s at the water-ice line. Due to the high fragmentation velocity, the grains in the outer regions of the disc grow larger and consequently drift inwards more quickly  compared to the case of low fragmentation velocities. However, inside the water-ice line, the low fragmentation velocity of the silicate grains reduces their Stokes number (see Fig.~\ref{fig:Stokes}), resulting in a slower inward drift. Consequently, the dust grains pile up inside the water-ice line,\footnote{This pile-up depends on the transition in fragmentation velocity. Without this transition, no pile-up is possible (see Appendix~\ref{ap:velocity}).} leading to an increase in the dust-to-gas ratio in the inner disc, independently of disc viscosity and disc radius. This can lead initially to dust-to-gas ratios higher than 0.3 in the inner disc, as estimated by \citet{2023arXiv230601653M}. However, in time the dust-to-gas ratio decreases significantly and even reaches  sub-solar values at the end of the disc's lifetime.

The inner disc is fed by the pebbles produced in the outer regions of the disc. As time progresses, pebbles form at ever larger radii (e.g. \citealt{2014A&A...572A.107L}) and drift inwards. However, once the outer edge of the disc is reached by this  `pebble-production line', the inward flux of pebbles decreases \citep{2014A&A...572A.107L, 2016A&A...591A..72I}. Consequently, the enrichment of the inner disc with small grains also decreases, resulting in a depletion of the dust-to-gas ratio, reaching sub-solar values after just 1 Myr of evolution. A larger disc radius allows a longer time to produce pebbles in the outer disc, resulting in a longer maintenance of a high pebble flux. Consequently, the dust-to-gas ratio depletes on a longer  timescale compared to smaller disc radii (top right in Fig.~\ref{fig:dtg}).

At higher viscosities the pebble sizes in the fragmentation limit are smaller (eq.~\ref{eq:frag}), resulting in a slower inward transport of pebbles. Consequently, the dust-to-gas ratio of the disc is initially reduced (bottom in Fig.~\ref{fig:dtg}) compared to simulations with higher viscosity. On the other hand, the slower inward drift of the outer pebbles  maintains the pebble flux for longer. In combination with the  slower inward drift of the small dust grains in the inner disc, higher   dust-to-gas ratios can be maintained for longer at high viscosity. Only towards the very end of the disc's lifetime does the dust-to-gas ratio in the inner disc reduce to sub-solar values (bottom right in Fig.~\ref{fig:dtg}).

The simulations   presented here show a significantly lower  dust-to-gas ratio compared to the estimate of \citet{2023arXiv230601653M}, especially after 0.5 Myr. The reason for this is the depletion of the pebble reservoir in the outer disc, compared to the assumption in \citet{2023arXiv230601653M} where a high pebble flux is maintained for the majority of the disc's lifetime. This effect is more pronounced for small discs than for large discs, meaning the dust-to-gas ratio reduces more quickly in smaller discs, in agreement with analytical estimates \citep{2016A&A...591A..72I}.

\begin{figure*}
 \centering
 \includegraphics[scale=0.47]{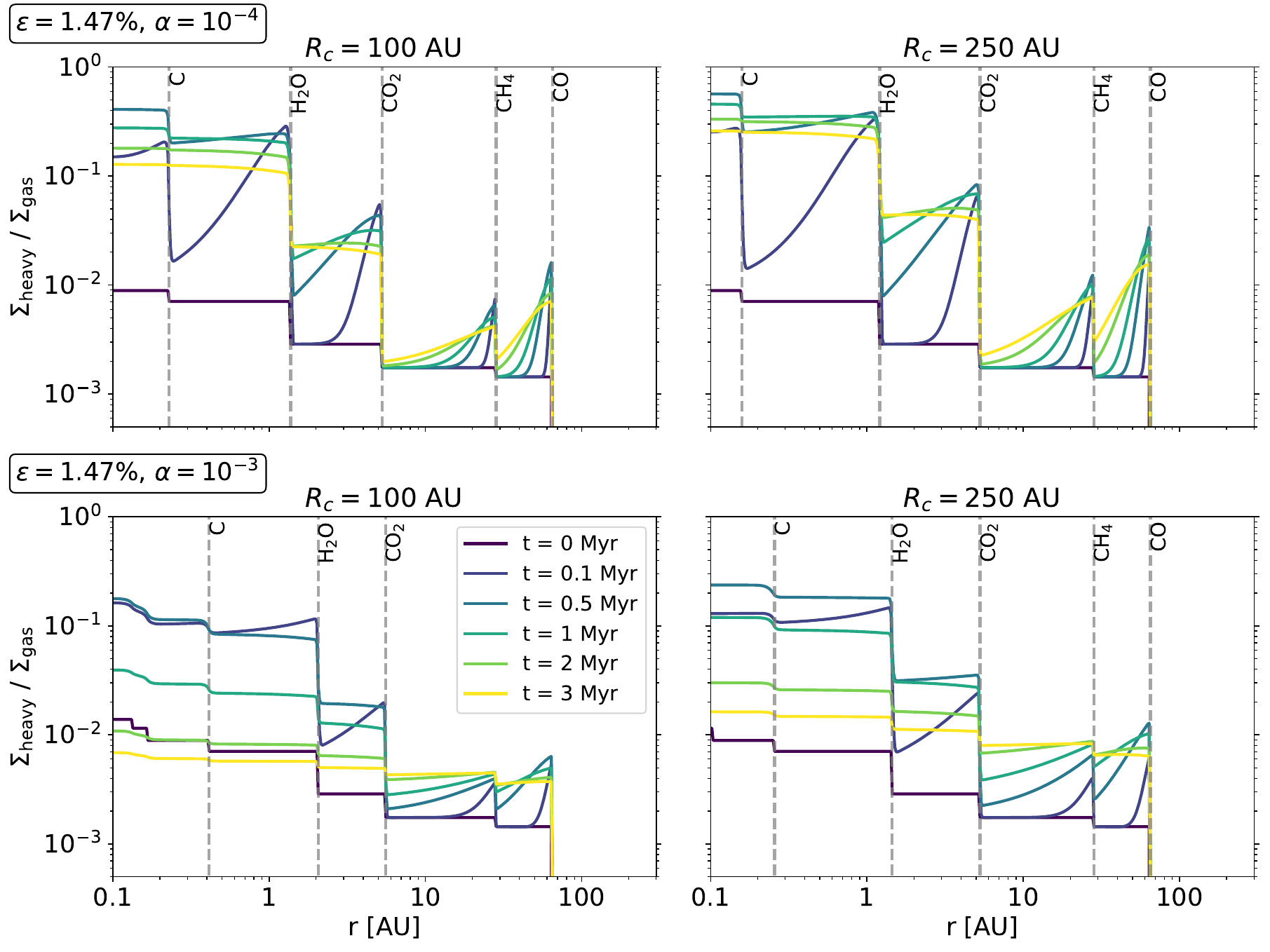}
 \caption{Time evolution of   heavy-element content in the gas phase in protoplanetary discs with $\alpha = 10^{-4}$ (top) and $\alpha=10^{-3}$ (bottom) and disc radii of $R_{\rm c}=100$AU (left) and $R_{\rm c}=250$AU (right). The vertical lines indicate the evaporation fronts of the different chemical species where inward-drifting pebbles evaporate, leading to increases in the heavy-element content of the gas phase. 
   \label{fig:heavy}
   }
\end{figure*}

Figure~\ref{fig:heavy} shows the heavy-element content of the gas phase ($\Sigma_{\rm heavy}$) produced by the evaporation of volatiles inside the evaporation fronts. This initially leads to very high heavy-element content just inside the various evaporation fronts. As time progresses, the volatile vapour is transported inwards by viscous transport, increasing the heavy-element content of the gas phase to largely super-solar values in the whole inner disc region  that can last the whole lifetime of the disc.

The duration of the vapour enrichment   depends on the disc radius and viscosity. The vapour enrichment is also fueled by the growing and inward-drifting pebbles originating from the outer disc. As soon as the pebble flux decreases, the vapour enrichment in the inner disc reduces on a viscous timescale \citep{2023arXiv230815128M}. Consequently, a disc with a larger radius that maintains the pebble flux for a longer time can keep its vapour enrichment for a longer time. This reduction is slower compared to the reduction of the dust-to-gas ratio because the vapour follows directly the evolution of the gas, while the small dust grains in the inner disc are slightly decoupled from the gas and move inwards faster than the gas. 

Discs with higher viscosity have smaller grains (eq.~\ref{eq:frag}), resulting in a slower enrichment of the inner disc regions with evaporated pebbles. Additionally, the higher viscosity results in faster vapour transport. Consequently, the vapour fraction in the inner disc reduces more quickly  over time once the pebble flux has decayed, and it cannot bring new material to the inner disc. This is exactly the opposite effect compared to the dust-to-gas ratio, where higher viscosities are beneficial to maintaining a high dust-to-gas ratio.

\begin{figure*}
 \centering
 \includegraphics[scale=0.47]{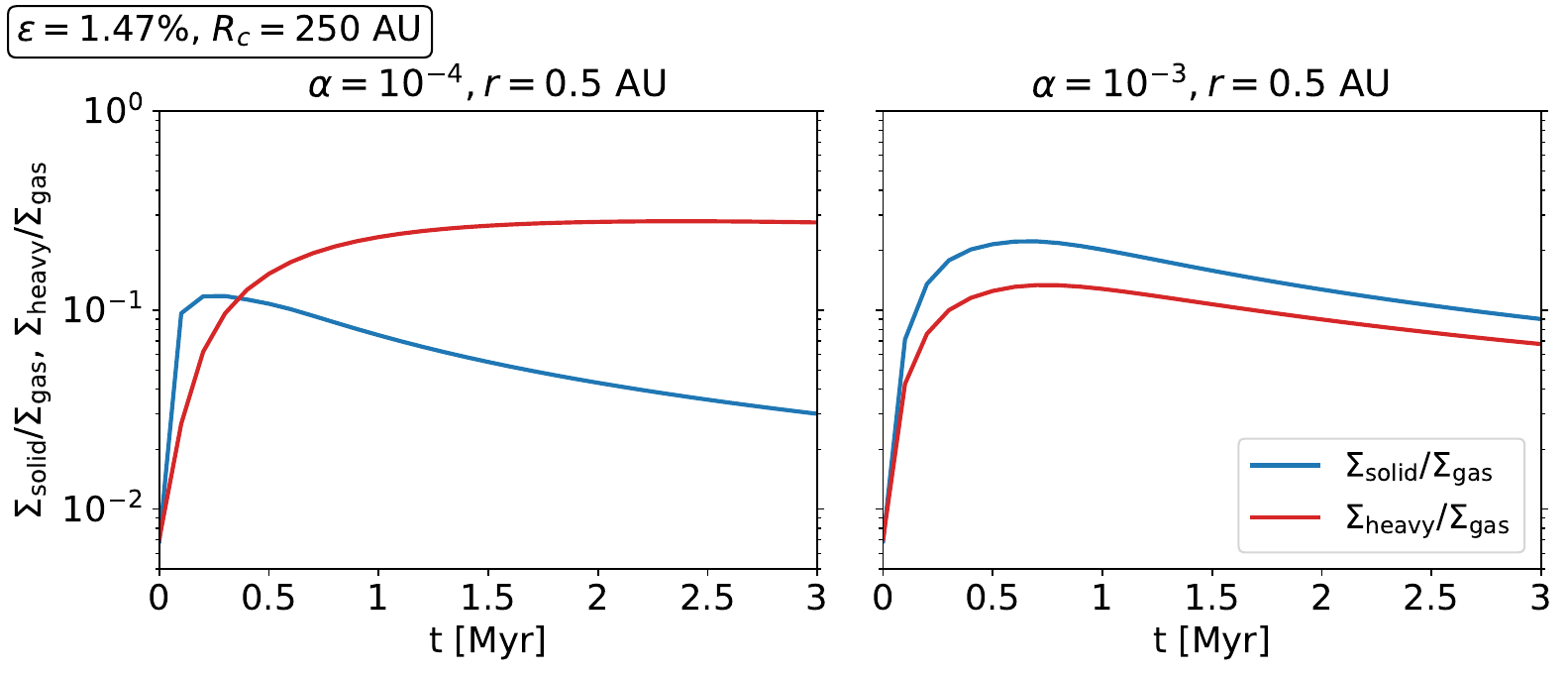}
 \caption{Time evolution integration of   heavy-element content of the gas phase at 0.5 AU either from the solids (see Fig.~\ref{fig:dtg}) or the vapour (see Fig.~\ref{fig:heavy}) for discs with $R_{\rm c}=250$ AU and $\alpha=10^{-4}$ (left) and $\alpha=10^{-3}$ (right).
   \label{fig:integral}
   }
\end{figure*}

Figure~\ref{fig:integral} shows the time integrated dust-to-gas ratio and heavy-element content at 0.5 AU, which is    between the water and carbon evaporation fronts in all our simulations. This time integration shows how the heavy-element content of an atmosphere of a putative planet that accretes with a constant rate would evolve. We only show the results for discs with $R_{\rm c}=250$ AU because these discs give the greatest heavy-element enrichment in dust (Fig.~\ref{fig:dtg}) and gas (Fig.~\ref{fig:heavy}). At low viscosity, the very initial enrichment by the solids is greater because of their faster inward drift, compared to the viscous transport of water vapour released at the water-ice line. After 0.3 Myr the heavy-element content in the gas phase is much higher compared to the dust-to-gas ratio in the inner regions of the disc. Especially at the end of the disc's lifetime at 3 Myr, the heavy-element content in the gas phase is about a factor of 10 higher than the dust-to-gas ratio.

On the other hand, at high viscosities, the picture reverses. The time integrated dust-to-gas ratio is higher than the heavy-element content in the vapour phase at all times. This is caused, as described above, by the smaller pebble sizes in the inner disc in combination with the faster viscous transport of the gas. However, in the end, the dust-to-gas ratio is slightly below 10\%, so a factor of 3 lower than anticipated by \citet{2023arXiv230601653M}.

The difference to \citet{2023arXiv230601653M} originates from a different disc model. \citet{2023arXiv230601653M} use the estimate from \citet{2016A&A...591A..72I}, relying on a constant $\dot{M}$ with radius. In this approximation, extending the disc to large radii results in a gravitationally unstable disc. In these environments, with Toomre Q$<$2, pebble growth is progressing only slowly, increasing the time until which pebbles are released from the outer disc \citep{2016A&A...591A..72I}. Consequently, a low pebble flux could be maintained for a long period of time. In contrast our here used disc model features an exponentail cut, $R_{\rm c}$, at either 100 or 250 AU, preventing $Q<2$ in the outer disc regions. Consequently, the grains in the outer disc can grow efficiently, resulting in a fast inward flux of pebbles and a fast decline of the pebble flux in our model. As a result, the enrichment of the inner disc with small dust grains can only be maintained for a short time (Fig.~\ref{fig:dtg}).

The heavy-element enrichment in the gas and dust phase depends on the overall metallicity of the system, where a higher metallicity allows a greater enrichment of the disc in vapour and small dust grains (Appendix~\ref{ap:metallicity}). Consequently, this can lead to higher heavy-element content in giant planets, as shown in \citet{2021A&A...654A..71S}.

\section{Heavy-element enrichment of planets}
\label{sec:planets}

To make an estimate of the maximum heavy-element content (from dust and/or evaporated pebbles) that a gas accreting planet could have, we follow the approach outlined in \citet{2023arXiv230601653M}, who argue that small dust grains can be successfully accreted with the gas flow onto a growing giant planet in the inner disc. The underlying idea is that   small pebbles follow the motion of the gas, effectively unhindered by the pressure perturbation of the planetary gap.

Morbidelli et al.  start with a 15 $M_{\rm E}$ core and then compute how the heavy-element content of the planet would change if it accreted gas enriched with small dust grains, as set by the local dust-to-gas ratio. In their model this ratio is 30\%. They thus argue that the 15 $M_{\rm E}$ core that accretes gas with this dust enrichment can reach a heavy-element content of around 100 $M_{\rm E}$ once it reaches the mass of Jupiter. However, as we show above (Fig.~\ref{fig:dtg}), the dust-to-gas ratio is actually much lower.

\begin{figure}
 \centering
 \includegraphics[scale=0.7]{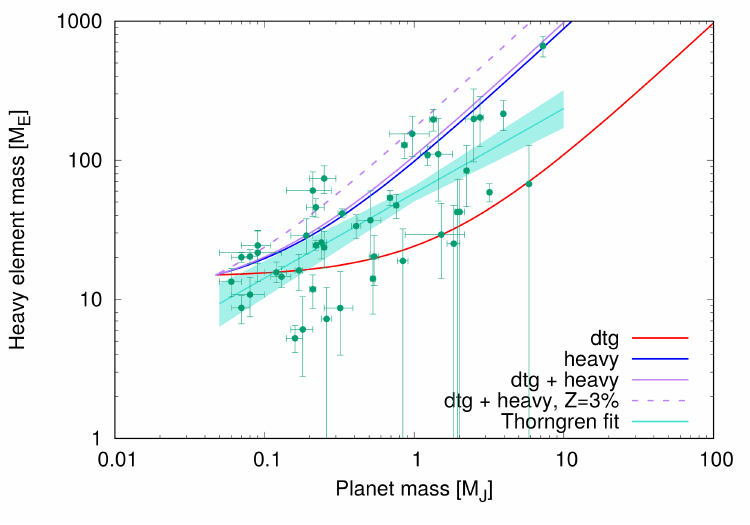}
 \caption{Heavy-element content of giant planets from \citet{2016ApJ...831...64T}. Plotted here is  the enrichment a 15 Earth mass core could have that grows by gas enriched with either dust (red) or vapour (blue). The purple line shows the combined heavy-element content of the planet if it accretes dust and vapour, while the dashed purple line shows the maximum enrichment a planet could have growing in a disc with a metallicity of 3\%.
   \label{fig:Thorngren}
   }
\end{figure}

Following their approach, we show in Fig.~\ref{fig:Thorngren} the heavy-element mass as a function of planetary masses for various estimates in our work, including  the constraints from \citet{2016ApJ...831...64T}. For our estimates of the total heavy-element content of giant planets, we use the assumption that the planet accreted enriched gas for 3 Myr in a disc with  with $R_{\rm c}=250$AU and $\alpha=10^{-4}$, corresponding to the final values on the left in Fig.~\ref{fig:integral}. Over the span of 3 Myr, the dust-to-gas ratio is around 3\%, resulting in only moderate heavy-element content that is too low to explain the data from \citet{2016ApJ...831...64T}. On the other hand, the heavy-element content of the gas phase results is 27.5\%, which  explains the majority of the data. 

However, the planet would be able to accrete small dust grains as well as vapour. Adding the heavy-element content of the gas phase and the dust component together, the planet can accrete gas with a heavy-element enrichment of around 30.5\%. If the overall metallicity of the disc were increase to 3\%, the total heavy-element content of small dust plus vapour is slightly above 50\% (see Appendix~\ref{ap:metallicity}). This  explains the large majority of the data, with the exception of a few planets with masses below 0.5 Jupiter masses, which might be very core rich and subject to heavy planetesimal bombardment \citep{2023arXiv230614740H}. However, the recent study by \citet{2023MNRAS.tmp.1812B} that additionally used constraints from atmospheric observations indicates that the heavy-element masses of objects in this mass range might be lower compared to the estimates by \citet{2016ApJ...831...64T}.

At higher viscosities, the time integration of the vapour content and the dust-to-gas ratio results in lower overall enrichment (Fig.~\ref{fig:integral}), which is why we do not display these results in Fig.~\ref{fig:Thorngren}. However, this illustrates a very important point: the final heavy-element content of a growing planet in this scenario is set by the disc parameters, and only a combination of vapour and dust accretion can explain the constraints derived from observations.

\section{Discussion and summary}
\label{sec:discussion}

In this section we discuss the caveats and implications of the  work presented here. In Appendix~\ref{ap:velocity} we show how the dust-to-gas ratio and the heavy-element content in the gas phase evolve in discs without any transition in fragmentation velocity at the water-ice line. The enrichment of the dust-to-gas ratio inside the water-ice line is fueled by the different particle sizes inside and outside the water-ice line, which result in different inward velocities that can generate a traffic jam in the inner disc region. The differences in the grain sizes could also lead to a pressure bump at the water-ice line, caused by an opacity transition  \citep{2021A&A...650A.185M}. However, without this velocity difference the  traffic jam does not exist, and thus no enrichment of the inner disc region can happen through this process. In contrast, the enrichment of the inner disc with vapour is independent of a transition in fragmentation velocity (Appendix~\ref{ap:velocity}) because the vapour enrichment is fueled by the relative velocity difference between the inward-drifting pebbles and the viscous transport of the gas. As long as the pebbles are much faster than the gas in the outer disc, the inner disc will be enriched in vapour (see also \citealt{2021A&A...654A..71S, 2023arXiv230701789K}). 

The motivation for the transition in fragmentation velocity originates from laboratory experiments \citep{2015ApJ...798...34G}, even though recent works have cast some doubt on this \citep{2019ApJ...873...58M}. Furthermore, chondrule sizes from the Solar System seem to indicate that the small silicate particles in the inner disc are mostly just millimetre-sized \citep{2015ChEG...75..419F}. However, VLA observations of inner discs show centimetre-sized particles all the way to the central star (e.g. \citealt{2020ApJ...890..130T}), which casts some doubt on the transition in fragmentation velocity. These regions are not fully resolved (e.g. \citealt{2020ApJ...890..130T}), however,  so in principle the very inner 1-2 AU could be devoid of these large particles. This issue can hopefully be solved with the ngVLA, which will provide more details on the structure of inner discs.

Planetesimals form from inward-drifting pebbles (e.g. \citealt{Johansen2014}). Recent recipes \citep{2017A&A...608A..92D, 2019ApJ...874...36L} have provided formalisms to calculate how the formation of planetesimals impact the flux of pebbles. Generally, as planetesimals form they take away pebbles from the disc, and thus reduce the pebble flux, which results in a lower enrichment of the disc with either small dust grains or vapour (Danti et al. in prep.). Here we are interested in the maximum value the effects of small dust grains and vapour enrichment can have on the heavy-element content of growing giants, motivating our choice to not include planetesimal formation in our model.

Our model relies on the assumption that the discs are smooth and that the pebbles can drift freely from the outer disc to the inner regions, in agreement with the observations of MP Mus \citep{2023A&A...673A..77R}. However, most discs show features, especially gaps and rings (e.g. \citealt{2018ApJ...869L..41A}), that indicate a trapping of pebbles in the outer disc. These traps, if efficient enough, prevent the efficient inward motion of pebbles (e.g. \citealt{2012A&A...538A.114P, 2021ApJ...915...62I, 2021ApJ...921...84K, 2021A&A...654A..71S, 2021A&A...649L...5B}), effectively preventing an enrichment of the inner disc with pebbles. Consequently, the dust-to-gas ratio in the inner disc will decrease. The heavy-element content in the gas phase will also decrease due to this effect (e.g. \citealt{2023arXiv230703846B}). However, volatile species that evaporate in the outer disc regions will still move inwards on the viscous timescale, allowing an enrichment of the disc \citep{2023arXiv230815128M}. On the other hand, the pebbles that are trapped in the outer regions can release small dust grains over timescales of several million years (e.g. \citealt{2023arXiv230105505S}), allowing a continuous supply of pebbles. However, by construction, this pebble flux will be lower than the flux in an unperturbed disc. Consequently the enrichment of the inner disc in dust and vapour will be lower compared to the  idealised case we presented here. We will investigate these processes in the future.

We assumed a very simplistic estimate to calculate the heavy-element content of giant planets. In reality, we expect the heavy-element content of the giants to be somewhat lower because the core of the planet has to form first, which might take a few hundred thousand years. However, during this time, the dust-to-gas ratio of the disc is already significantly depleted (Fig.~\ref{fig:dtg}), reducing the heavy-element content a giant planet could have; in contrast, the vapour content might not be greatly affected  (Fig.~\ref{fig:heavy}). We note that the detailed planet formation simulations based on the  pebble drift and evaporation of \citet{2021A&A...654A..71S} reach similar levels of heavy-element enrichment for planets above 2 Jupiter masses as our simple estimate, justifying our approach. 

In addition, as a giant planet grows it will open a gap in the disc, where the inward-drifting pebbles will accumulate. Small dust grains can diffuse through the gap \citep{2018ApJ...854..153W, 2018arXiv180102341B, 2018A&A...615A.110A}, where over timescales of several Myr the whole dust component could diffuse through \citep{2023arXiv230105505S}, provided the pressure trap can stop the large pebbles efficiently. In contrast, if the dust grains are very small, they are not hindered by the pressure bump and move inwards to be accreted by the planet effectively, as assumed by \citet{2023arXiv230601653M}. However, it is still unclear whether all the small dust grains in the pressure bump can actually be accreted by the giant planet, especially when considering the 3D geometry of the problem \citep{2018A&A...616A.116P, 2021ApJ...912..107B, 2021MNRAS.506.5969B}. The exact amount of small dust that can thus be accreted by growing giant planets is therefore not well constrained, and future simulations need to investigate this issue in detail.

The ratios of the vapour component to the dust-to-gas ratio could be used to estimate the volatile-to-refractory ratio of planetary atmospheres, which is thought to be a tracer of planet formation histories \citep{2021A&A...654A..72S, 2021ApJ...909...40T, 2022arXiv221109080C, 2022A&A...665L...5K}, even though the current measurements still show large error bars (e.g. \citealt{2019ApJ...887L..20W}). Our model predicts that planets with high total heavy-element content (30\% and above) should have a high volatile-to-refractory ratio in agreement with detailed planet formation simulations \citep{2021A&A...654A..71S, 2021A&A...654A..72S}. However, planets with a lower heavy-element content (below 20\%) could have volatile-to-refractory ratios of the order of unity (and below). This indicates that constraining the different planet formation pathways also requires   detailed knowledge of the heavy-element content of the giants compared to just its atmospheric abundances.

The enrichment of the inner disc depends on the overall composition of the materials. In our model, the water mass fraction of all the heavy elements is around 50\%, indicating that a large amount of water vapour can be released into the inner disc. However, if the stellar C/O ratio is higher, the water content generally is lower \citep{2020A&A...633A..10B, 2023arXiv230105034C}, reducing the amount of vapour enrichment in the inner disc. Additionally, low-metallicity stars ([Fe/H]<0) are enhanced in Mg, Si, O, and C compared to iron (e.g. \citealt{1957RvMP...29..547B, 2021MNRAS.506..150B}), indicating that more refractory materials that form small dust grains in the inner disc could actually be available, influencing the dust-to-gas ratio of the inner discs. Consequently, we suggest that the origin of the heavy-element content of giant planets should be re-examined on an individual basis rather than just with the iron component.

To summarise, we presented here a pathway that can allow the enrichment of giant planets with heavy elements, where the origin of these heavy elements, either from dust or vapour, depends on the disc parameters. In particular we identify that high enrichments of planets with vapour are possible for large disc radii and low viscosities independently of a transition in fragmentation velocity. On the other hand, a high enrichment  with dust grains also  requires, in addition to a large disc radius and a high viscosity,  a transition in fragmentation velocity at the water-ice line, which is still under debate (e.g. \citealt{2019ApJ...873...58M}). Due to the large variety in possible disc configurations, we naturally expect a diversity in the heavy-element enrichment of exoplanets.

\begin{acknowledgements}

B.B. and J. M. acknowledge the support of the DFG priority program SPP 1992 “Exploring the Diversity of Extrasolar Planets (BI 1880/3-1). B.B., also acknowledges the support of the European Research Council (ERC Starting Grant 757448-PAMDORA). We thank A. Morbidelli for discussions related to the different disc models.

\end{acknowledgements}

\bibliographystyle{aa}
\bibliography{Stellar}

\newpage

\appendix

\section{Model parameters}
\label{ap:parameters}

We describe here the parameters of our models that are not discussed in the main paper. For all the discs we use a fixed initial mass of 0.1 solar masses. Our initial gas surface density profile follows the analytical viscous disc solution of \citet{1974MNRAS.168..603L}, resulting in a surface density gradient of $\approx-1.08$. In time, the gas surface density of all chemical species is evolved via the standard viscous evolution equations using an $\alpha$-viscosity prescription.

As we use different disc radii, this implies that the gas surface density is lower for discs with larger disc radii. The disc's critical radius $R_{\rm c}$ is defined as the position of the disc, where the exponential decay of the gas surface density starts. The disc's temperature is determined through the equilibrium between viscous and stellar heating with radiative cooling. The temperature of the disc is fixed to the initial temperature and, for simplicity, does not evolve over  time. This implies that the positions of the evaporation fronts are fixed in time and only depend on the local gas surface density and viscosity.

For our model we use the solar elemental abundances \citep{2009ARA&A..47..481A}, where we take only the major elements into account. For simplicity, we do not use the more detailed chemical partitioning model, as in \citet{2021A&A...654A..71S}, but rather use the simpler model   presented in \citet{2023A&A...673A..17M} and shown in Table~\ref{tab:chempartition}. Using only these elements, we can calculate the self-consistent dust-to-gas ratio by summing over their mass fraction, resulting in a dust-to-gas ratio of 1.47\%.

\begin{table}
\centering
    \caption{Condensation temperatures \citep{2003ApJ...591.1220L} and volume mixing ratios of chemical species included in our model.}
    \label{tab:chempartition}  
    \begin{tabular}{c c c}
    \hline\hline
    Species & $T_{\text{cond}}$~(K) & volume mixing ratio \\ \hline
    CO            & 20  & 0.2 $\times$ C/H  \\
    CH$_4$        & 30  & 0.1 $\times$ C/H  \\
    CO$_2$        & 70  & 0.1  $\times$ C/H \\
    H$_2$O        & 150 & O/H - (3 $\times$ MgSiO$_3$/H + \\
           &      & 4 $\times$ Mg$_2$SiO$_4$/H + CO/H + \\
           &      & 2 $\times$ CO$_2$/H) \\
    C             & 631 & 0.6 $\times$ C/H \\
    Mg$_2$SiO$_4$ & 1354 & Mg/H - Si/H \\
    Fe            & 1357 & Fe/H \\
    MgSiO$_3$     & 1500 & Mg/H - 2 $\times$ (Mg/H - Si/H) \\ \hline
    \end{tabular}
    \begin{tablenotes}
     \item All material with evaporation temperatures of $T_{\rm cond} \leq 150$K are considered volatiles, while material with $T_{\rm cond} > 150$K are referred to as refractories.
    \end{tablenotes}

\end{table}

\begin{figure}
 \centering
 \includegraphics[scale=0.5]{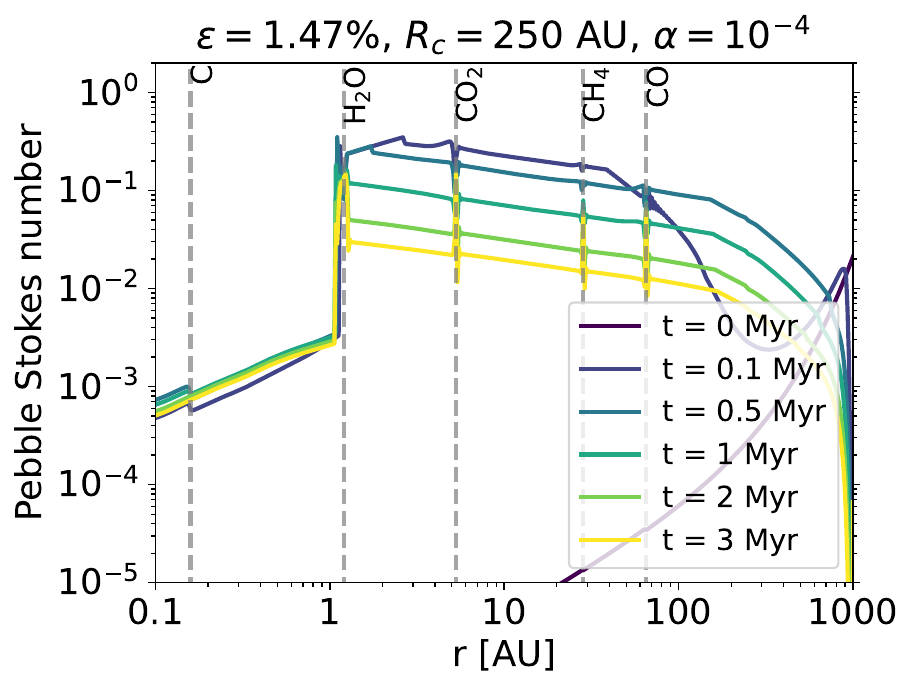}
 \caption{Time evolution of the Stokes numbers in our simulations with $R_{\rm c}=250$AU and $\alpha=10^{-4}$ and a transition of 1 to 10 m/s for the fragmentation velocity at the water-ice line position.
   \label{fig:Stokes}
   }
\end{figure}

Starting from the sub-micrometre-sized grains, our code calculates the growth and drift of the pebbles as a  function of time. In Fig.~\ref{fig:Stokes} we show the time evolution of the Stokes number of the pebbles in our simulations with a transition of fragmentation velocity of 1 to 10 m/s at the water-ice line. The physical size of the particles (eq.~\ref{eq:frag}) relates to the Stokes number as
\begin{equation}
 St = \frac{\pi}{2} \frac{a \rho_{\rm s}}{\Sigma_{\rm g}} \ ,
\end{equation}
where $a$ is the particle size. As we start the simulations with the same physical dust grain size, their Stokes number is initially radially increasing due to the decreasing gas surface density. After 0.1 Myr we observe a drop in the Stokes number at distances larger than 100 AU. This is caused by the long growth timescales in the outer disc regions, where pebbles have not fully formed at this stage. Due to the change in the fragmentation velocity by a factor of 10, we observe a change in the Stokes number by a factor of 100 around the water-ice line ($u_{\rm f}$ scales quadratically in eq.~\ref{eq:frag}). The resulting Stokes numbers of our simulations are in line with the disc model of \citet{2022A&A...666A..19B} used by \citet{2023arXiv230601653M}, allowing an easy comparison. In particular, our disc model allows a continuous flow of pebbles from the outer disc without any pebble traps that could hinder the inward drift of pebbles (e.g. \citealt{2012A&A...538A.114P, 2021ApJ...915...62I, 2021ApJ...921...84K}), as in the assumptions of \citet{2023arXiv230601653M}, even though the grain size differences at the water-ice line could lead to pressure bumps caused by an opacity transition \citep{2021A&A...650A.185M}.

\section{High metallicity}
\label{ap:metallicity}

\begin{figure*}
 \centering
 \includegraphics[scale=0.47]{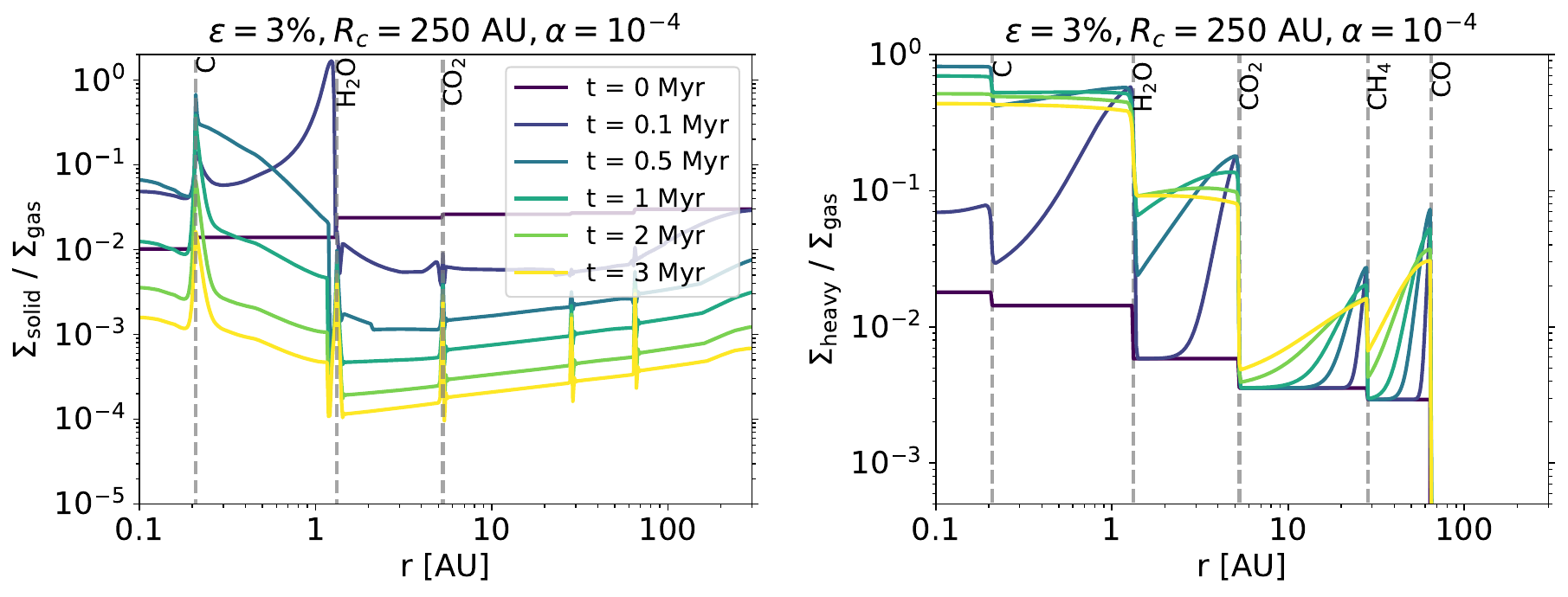}
 \caption{Time evolution of the dust-to-gas ratio (left) and the heavy-element content in the gas phase (right) for discs with $\alpha=10^{-4}$ and $R_{\rm c}=250$ AU. The disc's metallicity is 3\%, so nearly twice as high as our nominal simulations.
   \label{fig:dtg3}
   }
\end{figure*}

\begin{figure*}
 \centering
 \includegraphics[scale=0.5]{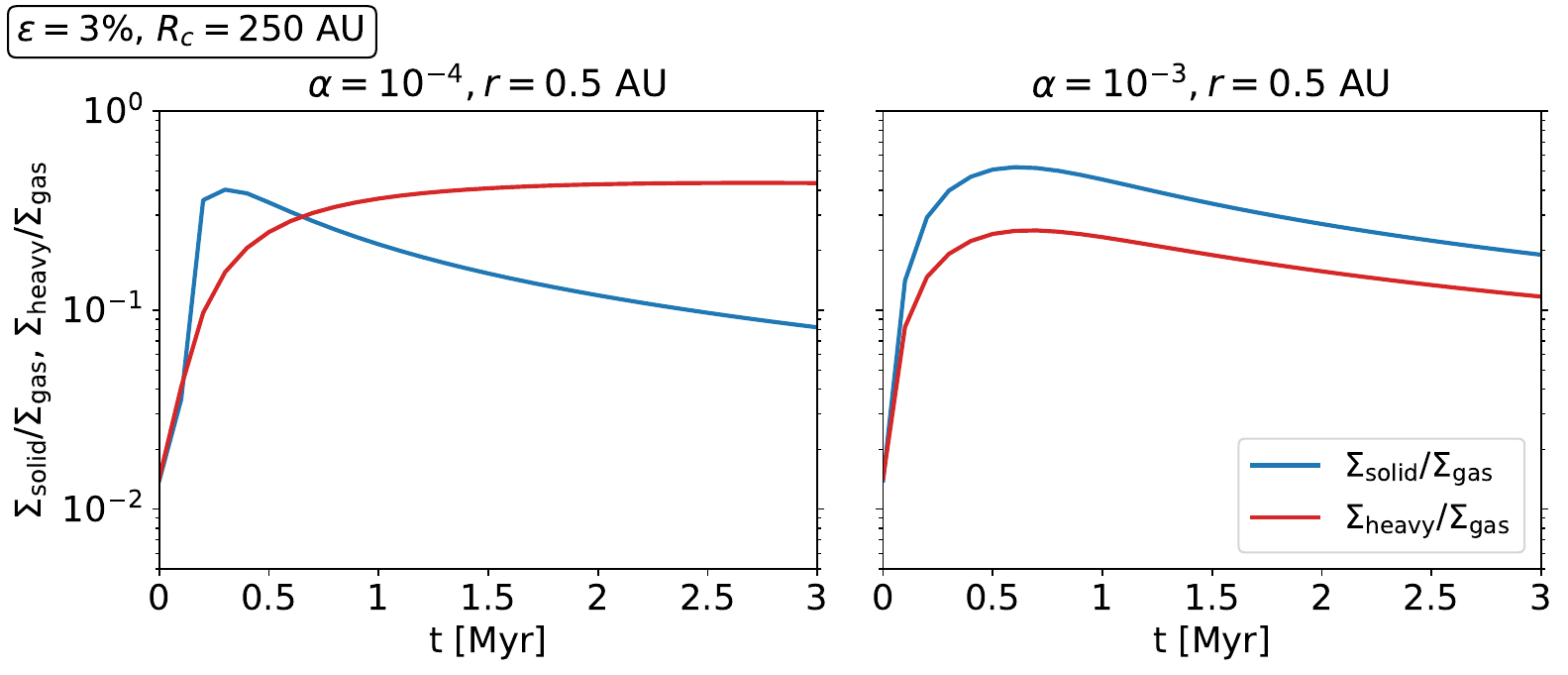}
 \caption{Time evolution integration of the heavy-element content of the gas phase at 0.5 AU either from   solids (see Fig.~\ref{fig:dtg}) or   vapour (see Fig.~\ref{fig:heavy}) for discs with $\alpha=10^{-4}$ (left) and $\alpha=10^{-3}$ (right). The disc's radius is $R_{\rm c}=250$AU and the   metallicity is 3\%.
   \label{fig:integral3}
   }
\end{figure*}

We show in Figs.~\ref{fig:dtg3} and  \ref{fig:integral3} the time evolution of the dust-to-gas ratio and of the heavy-element content in the gas phase, as well as the time integration at 0.5 AU for discs with a metallicity of 3\%. Considering our solar dust-to-gas ratio of 1.47\% ([Fe/H]=0.0), a metallicity of 3\% corresponds to [Fe/H]=0.31, a value slightly higher than the typical metallicities of hot Jupiters (e.g. \citealt{2018arXiv180206794B}).

As expected, the higher metallicity allows  greater enrichment in the dust-to-gas ratio and in the heavy-element content of the gas phase due to the larger amount of available pebbles that can drift inwards. However, this higher metallicity does not prolong the duration of the pebble flux, as also shown in the analytical estimates by \citet{2016A&A...591A..72I}. Therefore, the decline in the dust-to-gas ratio proceeds on  timescales similar to those for lower metallicity (see Fig.~\ref{fig:dtg3}). Nevertheless, the time integration of the dust-to-gas ratio shows higher values compared to the lower metallicity case (Fig.~\ref{fig:integral}).

The heavy-element content in the gas phase shows a similar behaviour. The larger amount of pebbles allows a greater enrichment of the inner disc in vapour, which is then consequently transported away by viscous transport. At low viscosity, the integrated heavy-element content of the gas phase stays above 40\% during the whole lifetime of the disc, while the dust-to-gas ratio drops   below 10\%.

At high viscosity, the dust-to-gas ratio is higher than the heavy-element content in the vapour phase, as for our standard metallicity. As expected, the total enrichment is also higher. This indicates that very high enrichment of planetary atmospheres are possible in a model that combines the accretion of evaporated pebbles and small dust grains, especially at high metallicity. This additionally shows the importance of detailed stellar abundance determinations for planet formation, as these set the refractory-to-volatile ratio.

\section{Constant fragmentation velocities}
\label{ap:velocity}

The fragmentation velocity of the dust grains sets their final size (eq.~\ref{eq:frag}). In particular a transition of the fragmentation velocity of 1 to 10 m/s at the water-ice line changes the grain sizes and their corresponding Stokes number (Fig.~\ref{fig:Stokes}) significantly. Some laboratory experiments indicate that the fragmentation velocities for silicate and water-ice grains might actually be very similar \citep{2019ApJ...873...58M}. Additionally, if bouncing during grain growth becomes important, the grain sizes might also be limited to similar sizes all over the disc \citep{2010A&A...513A..57Z}. Consequently, we investigate how the dust-to-gas ratio and the heavy-element content in the gas phase change if the fragmentation velocity of the grains is constant with respect to  composition, and thus to orbital distance.

\begin{figure*}
 \centering
 \includegraphics[scale=0.47]{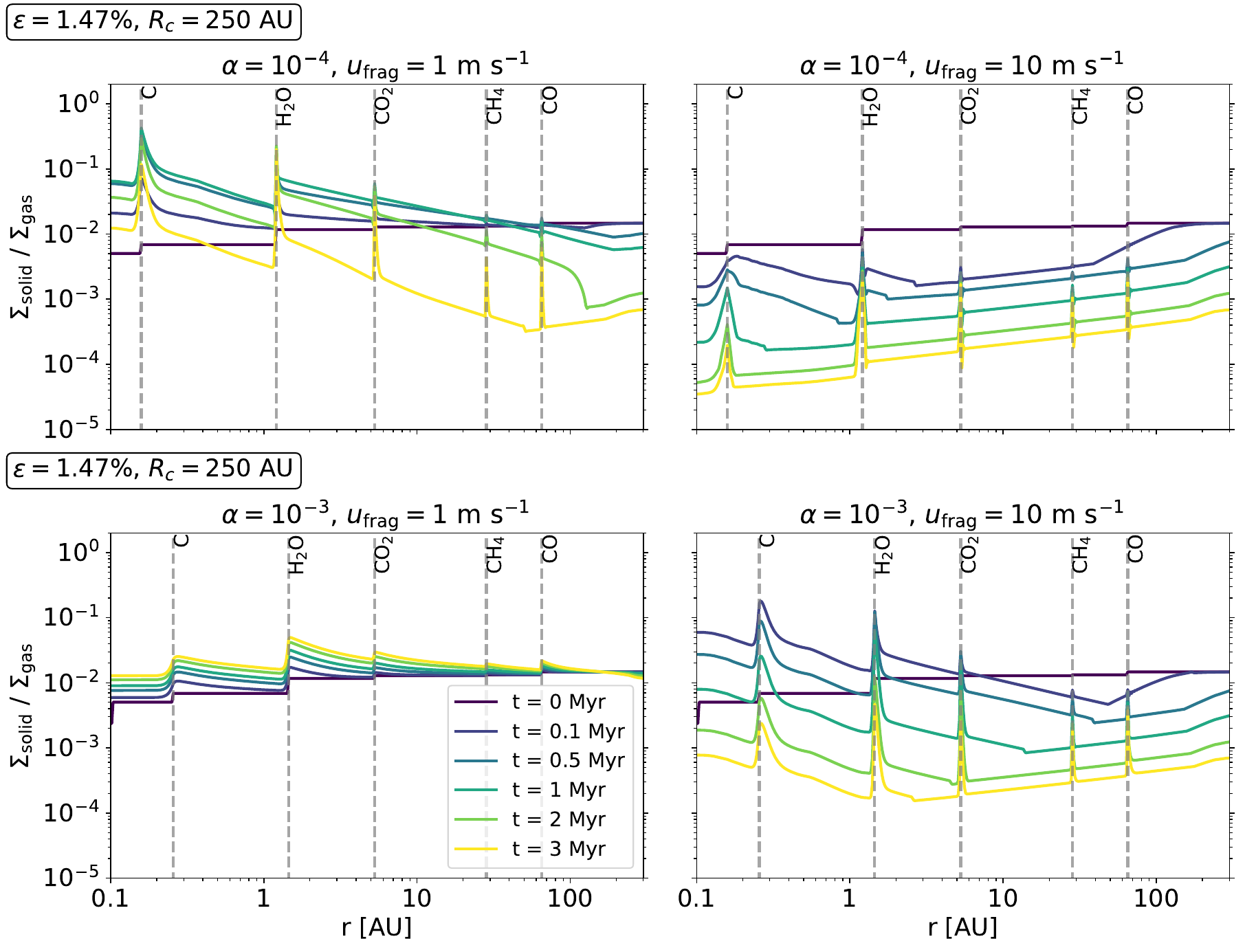}
 \caption{Time evolution of the dust-to-gas ratio in protoplanetary discs with $\alpha = 10^{-4}$ (top) and $\alpha=10^{-3}$ (bottom) and disc radii of $R_{\rm c}=250$AU. The pebbles in these discs experience constant fragmentation velocities of 1m/s (left) and 10m/s (right).
   \label{fig:dtg_uf}
   }
\end{figure*}

In Fig.~\ref{fig:dtg_uf} we show the solid-to-gas ratio in discs with $R_{\rm c}=250$ AU and different $\alpha$ viscosities for $u_{\rm f}=1$m/s (top) and $u_{\rm f}=10$m/s (bottom). As the grains grow, they drift inwards, release their volatiles at evaporation fronts, and continue as smaller dust grains inside the water-ice line. The pile-up of dust just inside the water-ice line is not observed because the particle sizes are continuous across the water-ice line. This prevents the pile-up of material compared to simulations with a transition in fragmentation velocity (Fig.~\ref{fig:dtg}). Consequently, the dust-to-gas ratios only reach values of up to 10\% in the inner disc regions for $u_{\rm f}=1$m/s and $\alpha=10^{-4}$. Higher fragmentation velocities allow the formation of larger grains that drift inwards very rapidly, resulting in a depletion of the dust in the disc on very short timescales, especially in the case of $u_{\rm f}=10$m/s and $\alpha=10^{-4}$, which harbours the largest pebbles. 

Higher viscosities show the same trend that a high enrichment of the dust-to-gas ratio are not possible for constant fragmentation velocities. The enrichment of the inner disc with dust is clearly a result of the traffic jam effect \citep{2017A&A...608A..92D}, where particles inside the ice line have smaller sizes compared to particles outside the ice line.

\begin{figure*}
 \centering
 \includegraphics[scale=0.47]{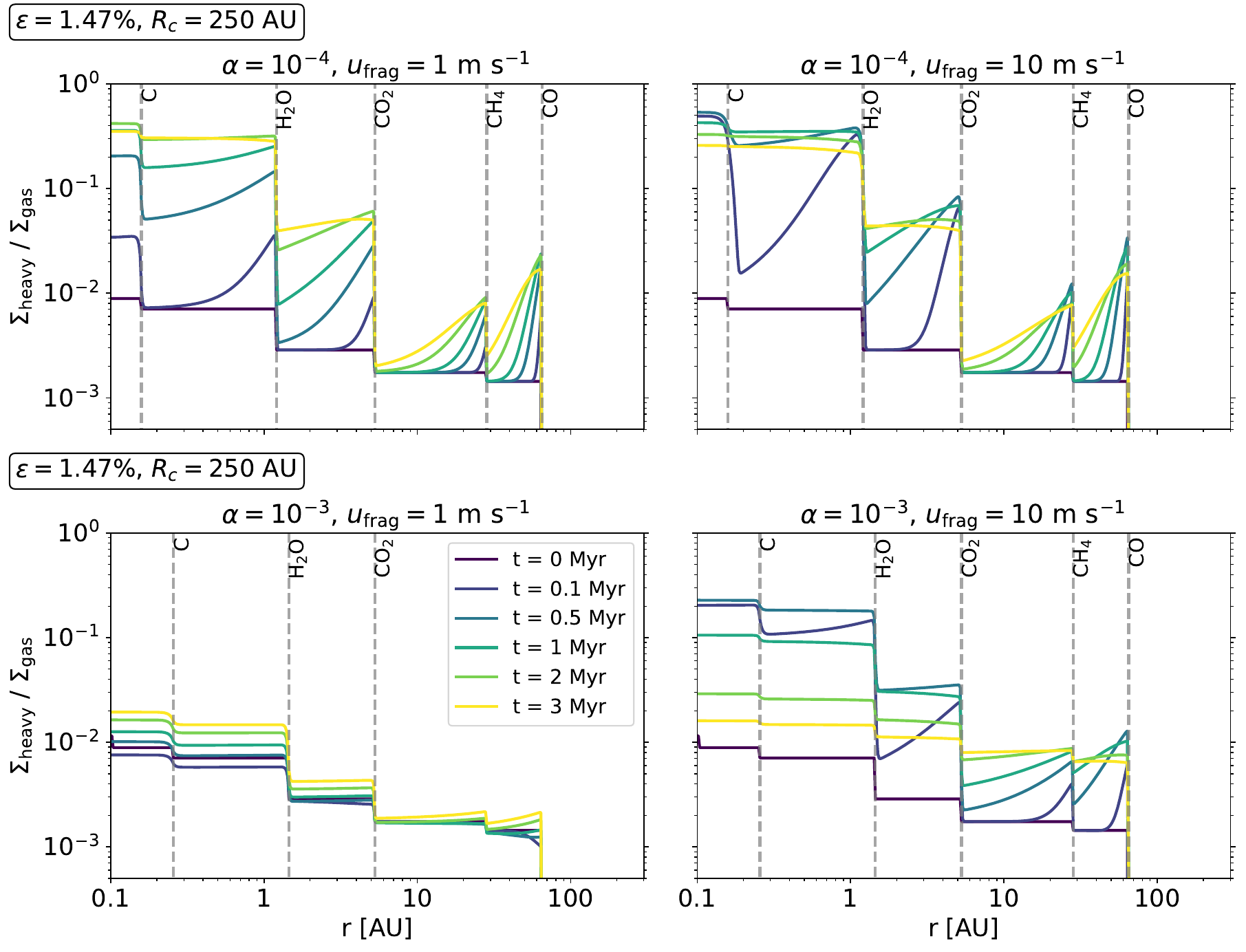}
 \caption{Time evolution of the heavy-element content in protoplanetary discs with $\alpha = 10^{-4}$ (top) and $\alpha=10^{-3}$ (bottom) and disc radii of $R_{\rm c}=250$AU. The pebbles in these discs experience constant fragmentation velocities of 1m/s (left) and 10m/s (right).
   \label{fig:heavy_uf}
   }
\end{figure*}

In Fig.~\ref{fig:heavy_uf} we show the heavy-element content in the gas phase for the same discs as in Fig.~\ref{fig:dtg_uf}. Inward-drifting pebbles evaporate and enrich the gas phase of the disc with their vapour. As for the simulations with transition in fragmentation velocities, the increase in the heavy-element content starts at the evaporation fronts where the volatile rich pebbles evaporate. Over  time, the vapour moves inwards, resulting in a nearly constant heavy-element content in the gas phase after a short period of time. The increase in the heavy-element content in the gas phase for the discs with a constant $u_{\rm f}=10$m/s is identical to the discs that undergo a transition in fragmentation velocity (Fig.~\ref{fig:heavy}). This is caused by the fact that the enrichment of the gas with vapour in the inner disc is driven by the pebbles that drift inwards from beyond the water-ice line and then evaporate. These pebbles have the same sizes in both sets of simulations, due to the same fragmentation velocities in these simulations.

This shows a very important difference between the two different approaches for the heavy-element enrichment. The enrichment of the dust-to-gas ratio in the inner disc region requires a transition in the fragmentation velocity, while the enrichment of the heavy-element content of the gas phase is independent of a transition in fragmentation velocity at the water-ice line. Of course, an enrichment that is sufficient to explain the heavy-element content of giant planets is only possible in discs that have a low viscosity because the high velocity transports the vapour away too quickly, leading to a low enrichment of the gas (e.g. bottom left in Fig.~\ref{fig:heavy_uf}).

\end{document}